# De- and Reconstructing Introductory Physics for the Life Sciences

Dawn C. Meredith[*] and Edward F. Redish[**]
[*] Department of Physics, University of New Hampshire, Durham NH
[**] Department of Physics, University of Maryland, College Park MD

**The Current State of Affairs: An opportunity**

For decades physics departments have been providing service courses for pre-medical students and biology majors. And for decades, physics faculty have been asking the question, "What do biologists and med schools want?" One of us (EFR) attended a conference discussing this topic in the summer of 1976.[1] The advice given there was rather generic, pointing to the wide diversity of life science classes, from "a course in Alaska for fish and wildlife managers to a course in Boston for biologists and a course in Kenya for agriculturalists. In each case the teacher is obligated to determine why, in content terms, those students are taking physics and to design the course accordingly."

Such an obligation, if seriously accepted, was challenge enough. But in the last few decades, the life sciences have experienced an explosive growth, presenting us with new challenges and opportunities. New probes, new instruments, and a growing understanding of the mechanism of life have enabled biologists to better understand the physiochemical processes that govern life at all scales, from the molecular to the ecological. Quantitative measurements and modeling are emerging as key tools for discovery. As a result, leading biologists are demanding new and more effective instruction at the undergraduate level, including improving the relevance of service classes in math, chemistry, and physics that could be the key to preparing students for the new, more quantitative biology. (See Sidebar 1.)

Biologists and pre-meds have traditionally been taught in separate course – "College (or algebra-based) Physics", as opposed to the "University (or calculus-based) Physics" recommended for physicists and engineers. At some institutions, College Physics has been dominated by biologists (UNH), but at many (UMd), College Physics serves a diverse collection of students ranging from pre-meds to architects to computer science and English majors needing a "lab science" class to meet a graduation requirement. This diverse audience has limited what could be done for biologists. But the recent explosive growth of biology not just as a science, but also as a science with an industrial base providing lots of good jobs has led to an explosive growth in the number of biology majors. At our institutions, we now teach physics to nearly as many biology majors as we do to engineers. This gives us an opportunity for rethinking whether we are serving them well, and for re-envisioning how we can provide opportunities for future biologists to appreciate the power of physics in helping them do the work they want to do.

We want our IPLS course to support biologists in the exciting work that they will do, yet many courses fall short of this goal. The stage for this mismatch was set decades ago, when the IPLS course likely originated (judging from older text books) by

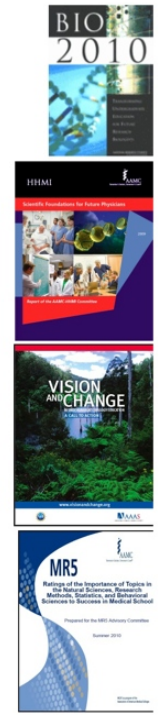

**Sidebar 1:**

Calls for change from biologists: Over the past decade there have been several calls for change to the undergraduate education of future biologists and medical professionals. These documents note that while the work of professionals has changed dramatically, their education has not. Of most interest to IPLS instructors is that the documents call for the education of life scientists to be more quantitative and interdisciplinary, and for the students to be more actively engaged in their education. Each document gives an insightful overview of the educational needs of their target cohort from the viewpoint of practicing professionals. We list below the sections of each report most relevant to IPLS instructors:

- Bio 2010: *Transforming Undergraduate Education for Future Research Biologists*[11]: Appendix E focuses on physics and engineering topics and the value of these disciplines to research biologists.

- *Scientific Foundations for Future Physicians:*[12] The entering medical school expectations E1-E3 are most relevant for the IPLS course. Importantly, the report suggests that medical college admission be based on competencies rather than courses taken; this allows colleges and schools to meet the competencies in many different ways.

- *Vision and Change in Undergraduate Biology Education, A call to Action.*[13] This document does not give any specific physics competencies, but does outline the core biological concepts and core skills for students of biology, as well as educational strategies to engage students more deeply and more authentic assessment practices. And it argues strongly that biology is inherently interdisciplinary.

- *MR5: Ratings of the Importance of Topics in the Natural Sciences, Research Methods, Statistics, and Behavioral Sciences to Success in Medical School.*[3] The AAMC has been reconsidering the requirements for med school and a new MCAT will be deployed in 2015. This discusses the most relevant physics content for pre-meds. An MCAT preview with sample questions is given in a recent report.[14]

taking the text for engineers and physical scientists and stripping out the calculus. An occasional biology application was added, but these would not be recognized as relevant by any biologist and did not address the real concerns of biologists. In addition, the choice of topics was unchanged from the calculus-based course, implying that all the topics (even heat engines!) are worth the necessary intense intellectual effort of biology students.

This lack of relevance is even more serious given that this is likely the last physics course our life science students will take; what we give them in two semesters is all they will get. If we leave them with the sense that physics is hard and of no use to a biologist, that perception is likely to remain with them for the rest of their lives. How can we expect a novice in both physics and biology to make connections that experts in either field would struggle to make, and when the topics they need to connect physics to biology are introduced months or years apart, with different terminology and different goals in very different contexts? A course without obvious relevance is understandably not well received by the students, and they are often advised by their biology advisor to wait to take physics until after they have taken "the important courses," – often in their junior or senior year. Biology instructors (probably correctly) then have to assume that not all students have had physics, so they do not make the connections in their classes either, confirming the view that physics is not needed for biology.

In recent years, spurred by numerous policy documents (see Sidebar 1) and a new focus in both biology and physics on education research,[2] the course at many institutions has improved (see Sidebar 2). Both of us have been involved in rethinking Introductory Physics for the Life Sciences (IPLS) for many years.[3,4,5] For both of us, this activity has brought with it a close interaction and collaboration with biologists. From this, we have learned that bringing a physics course into alignment with the needs of biology students was a subtler and more complex activity than we had imagined. Certainly, there needs to be more examples drawn from the life sciences. Certainly, there needs to be some shifts in the content covered by the class. But what we have learned is that both of these shifts are more serious than we had first expected.

Examples need to not only be drawn from a life

---

**Sidebar 2**
There has been much activity nationwide to improve the Introductory Physics for Life Scientists (IPLS) course. Many of these resources are publically available:

- IPLS sessions at AAPT national meetings have been held twice yearly for the past four years, and are likely to continue in the future.
- A group of IPLS educators has created a wiki with several resources for other instructors, as well as links to other sites: http://www.phys.gwu.edu/iplswiki/index.php/Main_Page
- A searchable database of educational resources for future physicians is available at https://www.mededportal.org/icollaborative
- Textbooks: it is beyond the scope of this article to give a comprehensive review of textbooks for the IPLS course, but we note that there are many textbooks that address the needs of biology students in much deeper ways than have been previously done. Two older textbooks are also worth a look: *General Physics* by Sternheim and Kane, and *Physics With Illustrative Examples From Medicine and Biology* by Benedek and Villars.[10]

---

science context, but to lead to authentic biological value for the students – a perception that they understand the biology they have learned more deeply as a result of learning physics. Content needs to change not only slightly, but potentially dramatically. It's not enough to skip relativity and put in a little fluid dynamics. The entire set of contextual assumptions we make in the class need to shift. And even more powerfully, we have learned that what the biologists bring to our classrooms is not just "less skill in mathematics than the engineering students on the average" but a deeply different perception of what science means and deeply different expectations of how to do it.

Physics has much to offer biology students, but what we traditionally offer them doesn't cut it. It is as though we are educating furniture makers by having them build houses – superficially there is some similarity, but really, it's not the best we can do. A serious rethinking of IPLS is required and has begun to stir groups of educators at AAPT and APS meetings (see Sidebar 2). This article is intended to open the debate to the broader physics community.

**Listening to biologists: Disciplinary differences**

The first step in rethinking IPLS to meet the needs of biologists is for physicists and biologist to better understand the way they each think about their science and the way they bring their science to undergraduates. For each of us, developing our understanding of biologists relied on many conversations with many biologists, but most importantly, with an extensive interaction with a single biologist over a period of many years. (For Dawn, this was Jessica Bolker, an ecologist at UNH.[3] For Joe, this was Todd Cooke, a botanist at UMd.[5]) We both found our conversations with biologists immensely valuable; not only in helping us understand biology and biology students, but also in developing deeper insights into our own understanding of physics and the many hidden assumptions that went into our own teaching of undergraduates.

One critical fact that emerged early and continually in our conversations with biologists is that there were dramatic cultural differences between physicists and biologists. This is not to say that there are not biologists who think and behave like physicists (some are physicists) or vice versa. But there does seem to be a "cultural average" or "mean field" that is different for the two professions that shows up strongly in discussions of the appropriate way to teach introductory science. This affects both the epistemological style of the class (what one takes as given and what one derives) and the content that each group sees as appropriate. Some of the differences we have found are articulated in the Sidebar 3.[5]

We note that while there is overlap in some aspects of physicists' and biologists' approaches to introductory science classes, they tend to see them differently. For example, some of the biological models used in introductory classes can be described as toy models – highly unrealistic and introduced for the purpose of understanding one component of a mechanism. The Hardy-Weinberg model of evolution is one such, relying on unrealistic assumptions. But this is not as common in biology as in physics.

Some of the biologists we have spoken to considered traditional toy-model physics examples, even such central and powerful ones as the simple harmonic oscillator (mass-on-a-spring), irrelevant, uninteresting, and useless until the physicists were able to show its value as a starting-point model for

---

**Sidebar 3**

Here are some of the differences that we often find lead to "interesting discussions" between physicists and biologists.

*Physics: Common epistemological framings*

P1. Introductory physics classes often stress *reasoning from a few fundamental (usually mathematically formulated) principles*.
P2. Physicists often stress building a complete understanding of the *simplest possible (often highly abstract) examples* – "toy models" – and often don't go beyond them at the introductory level.
P3. Physicists *quantify* their view of the physical world, *model with math*, and *think with equations*, qualitatively as well as quantitatively.
P4. Physicists concern themselves with *constraints* that hold no matter what the internal details (conservation laws, center of mass, …).

These elements will be familiar to anyone who has ever taught introductory physics. What is striking is that we rarely articulate this for students – and none of these elements are typically present in an introductory biology class. Biologists have other concerns.

*Biology: Common epistemological framings*

B1. Biology is often *incredibly complex*. Many biological processes involve the interactions of component parts leading to emergent phenomena which includes the property of life itself.
B2. Most introductory *biology does not emphasize quantitative reasoning* and problem solving to the extent that it serves in introductory physics.
B3. Biology contains a critical *historical constraint* in that natural selection can only act on pre-existing molecules, cells, and organisms for generating new solutions.
B4. Much of introductory biology is *descriptive* (and introduces a large vocabulary)
B5. Biologists (both professionals and students) focus on and value *real examples* and *structure-function relationships*.

---

many real-world and relevant biological examples. This required making it clear from the first that a Hooke's law oscillator was an *oversimplified model* and illustrating how it would be modified for realistic cases. This is unfortunately rarely done in introductory physics classes and physicists are typically taken to task for "wanting to live in frictionless vacuums." (See Figure 1.)

Figure 1.   (from XKCD)

*http://xkcd.com/669/*

**What content should be covered?**

These different epistemological inclinations lead to different ideas as to what needs to be emphasized about content. In addition, physics instructors have become so comfortable with the traditional physics majors and engineering-based introductory physics course that the decisions that are made in constructing that course have become invisible and seem "just the way things have to be done." When told about our plans to renegotiate the content of IPLS, one of our physics colleagues told us,

> *I would be inclined to approach it from the "other end": i.e., I would construct a list which has in it the absolute irreducible physics concepts and laws that have to be in a physics curriculum. This "entitlement" list will already take up a majority of the available space.*

Once we get into detailed negotiations with biologists, it becomes clear that these "absolute irreducible physics concepts" are not absolute, are not nearly complete ("have to be in a physics curriculum"), and have been selected based upon unstated – and unrealistic – assumptions about who the students were and what they would need to bring into their next physics class. The fact that, at present, most biology students will have no "next physics class" (though we hope this will change in the future) changes dramatically what is "entitled."

Some of our physics colleagues complain that a course with a different list of "irreducible physics concepts and laws" that those in the course for physicists and engineers is not a real physics course. While we agree that in an ideal world all students would learn to love physics and see it as we do, this is simply not a likely outcome of a one-year course, for students who have their interests elsewhere. We argue that the course that we propose has the possibility of giving students an appreciation for physics, and a working knowledge that will serve them well later. We believe that this seemingly modest goal is both difficult to achieve and worthwhile.

As one example of changes to what is entitled, consider that strong emphasis that traditional introductory physics places on point masses and rigid bodies. There are good reasons for doing this. Ignoring an object's extent (treating it as a point mass) allows us to focus on one variable kinematics and dynamics. It allows us to build the concepts of inertia, interactions, forces, and energies with mathematical clarity and precision. (Epistemological framings P1 and P2.) It is of course a false precision. No object is a point and for realistic objects, much of their real world behavior depends on their structure and extent. Biologists are particularly sensitive to this (Epistemological framing B5) and are highly suspicious of ignoring structure – which often has critical implications for molecules, cells, and organisms.

When we relax the point-mass approximation in introductory physics, we tend to assume objects are rigid bodies. Again, this allows us to move to more complex situations with mathematical clarity and precision. We know there are no such things as "rigid bodies" but we are content to ignore deformations at speeds small compared to the speed of sound in the object and to suppress deformations in favor of introducing the phenomenological idea of "normal forces" in which the deformation associated with it is (usually) hidden. [Some modern textbooks model normal forces as spring-like.] Rigid or nearly rigid bodies are common in mechanical engineering, but are rarely a good approximation in biology. Yes, there are trees and bones, but in cellular and sub-cellular biology, which is coming to dominate introductory biology instruction, nothing can be remotely approximated as rigid.

A second example that illustrates our hidden assumptions is our isolation of objects. In line with epistemological framing P2, we tend to isolate our objects so that we can focus on fundamental processes with a small number of objects and interactions. In biology, however, essentially everything takes place in a fluid environment – either air or water – that is not "peripheral" or ignorable but that plays a critical role in the functioning of an organism. We are both theoretical nuclear physicists. Nuclear physics is a place where the environment – in most cases even the electron cloud bathing the nucleus – can be completely ignored in all but the most sensitive measurements. [Some with biological relevance – such as NMR!] The idea that a chemical reaction rate could be affected by pH seemed startling to at least one of us. This suggests that the behavior of fluids and the behavior of matter imbedded in fluids is, for biology students, a part of the "absolutely irreducible physics concepts and laws" that need to be included in an introductory physics class, and that tacking on a discussion of Bernoulli's principle or the Hagen-Poiseuille equation (Ohm's law for fluid flow in a pipe) will not suffice.

A third example is projectile motion and the inclined plane. These are essentials in every introductory physics class. Why? Well, many mechanical engineers will encounter projectiles and inclined planes help one learn about mechanical advantage – neither of which is of much use to most biologists, except in rather specialized circumstances – such as the biomechanics of skeletal muscle systems. For many physicists, though, they are among the cleanest examples possible of epistemological framings P1, P2, and P3. They have surprising results that show the power of mathematical principle-based reasoning and both are places where students can develop the basic skills of vector analysis. Having made these identifications, we can, perhaps, find other contexts that achieve the same goals but that also seem to have biological authenticity.

Another way to answer this question about what should be covered is to ask the biologists. A committee of the Association of American Medical Colleges has addressed this issue,[6] and at several institutions, including our own, we ourselves have made this inquiry with the hope of narrowing down the topics that need to be included in the IPLS course. Invariably, we find that there is at least one biologist that is interested in almost every physics topic typically covered in an introductory course. This problem arises because biologists themselves are not all cut from the same cloth, and some content needs are specific to sub-disciplines of biology. For example, those that work at the cellular level need to understand entropy because it is the free energy (*E-TS*) that is a measure of how much useful work can be done. For those biologists that work at the organismal level, forces are important: physical therapists are concerned about torques that arise as limbs move, and organisms must push on the environment to move. Those that study populations and eco-systems need to know about feedback loops and system dynamics. Unfortunately, some biologists downplay the needs of those in other sub-disciplines, so we must be the champions for all of our students' needs.

Even with this wide variety of needs, there are some topics that we agree should be de-emphasized including projectile motion, rotations with constant acceleration, Newton's law of gravitation, and heat engines. Other topics gain new prominence: fluids, optics, energy, and entropy. And still other topics that are not mentioned at all in the standard algebra-based course should be considered for inclusion: scaling, strength of materials, gradient-driven flows, and descriptions at the molecular and macroscopic scale. In the end, we are left with the advice in the opening paragraph: know your students and what physics is likely to be useful to them. If our population is primarily kinesiologists, physiologists, and pre-physical therapists our course ought to include lots of discussion of forces, motion – even projectile motion, and deformations of solids. But as the population of biologists increases, for this kind of specialty to dominate a population of biologists is increasingly rare. Strong biology programs increasingly are focusing their students' attentions on biochemistry, genetics, molecular biology, and cell biology.

What physics should be taught for this growing population of future biologists? Topics should be chosen because they have authentic biological applications but also because they contribute to a coherent and cohesive story line for the physics – and help to develop and strengthen those scientific skills and competencies that are most effectively taught in a physics context.

**What process skills and competencies should we teach?**

There is far more agreement among physicists and biologists about the scientific process and competency goals for an IPLS course than there is about content. Foremost, physics provides a wonderful context in which students can learn to synergistically blend quantitative work with sense-making. Because of its emphasis on highly simplified models (Einstein's razor – "as simple as possible but not simpler"), physics is an excellent context for helping students develop skills of scientific modeling, especially with math, developing strong physical pictures and mechanism integrated with multiple representations, and developing a sense of the quantitative. As long-time physics instructors, we feel confident that we know how to do this. But to develop a course that not only satisfies the instructor but also meets the needs of biology students for biological authenticity can be more challenging.

One of us (EFR) has been working with a biologist who has been bringing more physics and more explicit mathematical reasoning into his biology class, Organismal Biology (OrgBio).[7] Interviewing the students in this class (and in physics) helps us understand some of the attitudes and expectations that biology students are bringing. In one activity in OrgBio, students had to work with Fick's second law of diffusion, $<x^2> = 4Dt$, to understand why certain biological structures (gills, lungs) developed. Here is one student's response.

> *I don't like to think of biology in terms of numbers and variables. I feel like that's what physics and calculus is for. So, I mean, come time for the exam, obviously I'm gonna look at those equations and figure them out and memorize them, but I just really don't like them. I think of it as it would happen in real life, like if you had a thick membrane and you try to put something through it, the thicker it is, obviously the slower it's gonna go through. But if you want me to think of it as this is x and that's d and then this is t, I can't do it. Like, it's just very unappealing to me.*

Later in the interview, she recalled a different activity using math and physics, and in this one she had a dramatically different reaction. The instructor was talking about scaling laws and the importance of surface to volume ratios by showing an example of a wooden horse made with blocks of wood for the head and body and dowels for the neck and legs. A small model stood just fine, but a model with every dimension scaled up by a factor of two broke the dowel legs. The explanation was that the mass increased with the cube of the dimensions but the support strength of the legs depended on the cross sectional area, which only grew as the square of the dimensions. The student got excited in describing this result.

> *The little one and the big one, I never actually fully understood why that was. I mean, I remember watching a Bill Nye episode [TV science program in the US] about that, like they built a big model of an ant and it couldn't even stand. But, I mean, visually I knew that it doesn't work when you make little things big, but I never had anyone explain to me that there's a mathematical relationship between that, and that was really helpful to just my general understanding of the world. It was, like, mindboggling.*

Interestingly enough, this student subsequently took our reformed[4] algebra-base physics and did very well in both terms. It was clear from her performance that her objections to the use of math were not based on a lack of ability to use math. What we learn from this and other interviews is that biology students often bring cultural/disciplinary expectations to their classes that may get in the way of trying to create interdisciplinary instruction – but they can be context dependent. If the students perceive the activity as doing work for them, they respond much better.

As we think about teaching process goals, we must acknowledge that, on average, biologists are less fond of and less adept at using mathematics than the average engineering student, and some students may even rebel against using mathematics to describe biology, as we have seen above.[8] Yet these students will be expected to use mathematics in many of their upper division biology courses, and we have an opportunity to make a real difference in their preparation to do this.

One of us (DM) found in discussion with a biology colleague that students from our course were coming into his upper division course much less terrified and more accepting of physics and mathematics as applied to biology; this is the sort of outcome that makes a significant difference in upper division biology courses. Therefore IPLS students should be given multiple opportunities to practice this unfamiliar art through estimates, scaling arguments, explorations of basic functions (exponential, logarithmic, trigonometric, polynomial), inferences from equations, modeling well-understood phenomena with equations, checks for coherence, and multiple representations of quantitative relationships (plots, equations, tables of values). But in order to catch their attention, we need to do this in contexts that biology students find meaningful.

There are many simple examples of each of these experiences with quantification and sense making. For example, one implication from the definition of kinetic energy is that KE can be minimized by minimizing the mass of the legs, the fastest moving part of the animal. Students can easily be led to infer the equation for pressure drag knowing that it is caused by collisions with fluid molecules. At UNH, we developed a cooling lab that allowed students to connect the parameters in Newton's law of cooling with physical quantities (initial temperature, room temperature, amount of insulation). A group of students who were videotaped in this activity showed intense engagement and negotiation of meaning. At the end of the lab, one student remarked, "It's kinda cool way to make us figure out this equation." And his lab partner allowed that it was.[2] Biology students are often not used to making sense of equations, and we have many opportunities to help them do this.

The one thing we believe they do not need is extended proofs: our experience and research show[9] that novices are hard-pressed to make any sense of proofs because they do not have the rich context that gives proofs their meaning. (One of us has even seen our junior level physics majors struggle to see the point of a proof.) While it is both useful and valuable for biology students to learn how to work with symbols and think with equations, extended derivations are likely to be a hard sell. It is probably not possible with classes of general biologists to be able to spend the time to develop the skills of extended mathematical derivations. (This of course no longer holds for explicitly interdisciplinary populations of students such as bioengineers, biophysicists, and mathematical biologists.)

> **Sidebar 5:**
> Gradient-driven flows: Potential across a membrane. This is an ideal situation (as we had in lab); it is NOT meant to describe a real cell membrane (though there are some similarities).
>
> (1) You have a membrane with pure water on the left side and $K^+$ ions on the right side. The membrane is permeable to $K^+$
>
> (a) What gradients, if any, are present in this situation?
>
> (b) Will the $K^+$ move across the membrane? If so, in what direction and why?
>
> (c) If there is net motion of $K^+$ when will it stop (if at all) and why? If there is no net motion, why not?
>
> (2) You have a membrane with $Na^+$ on the left side and $K^+$ ions on the right side (equal numbers of ions on both sides). The membrane is permeable to $K^+$ only.
>
> (a) What gradients, if any, are present in this situation?
>
> (b) Will the $K^+$ move across the membrane? If so, in what direction and why? Is there a net motion of $K^+$?
>
> (c) If there is a net motion of $K^+$ when will it stop (if at all) and why? If there is no motion, why not?

**Problem Solving for Biologists**

The overarching content question that life science students need to investigate is, "How does the physical world both constrain and facilitate the work that molecules, cells, and organisms must do?" More specific questions include, "How do organisms solve the problems of gathering, storing, and efficiently using energy?" "How do organisms create organization and retain information?" "How is structure related to function?" "Why do animals behave as they do?" "How do organisms communicate efficiently?" The kinds of problems that our biology students work on address these issues. Our students should also be able to build simple quantitative models of biological systems, know what physics matters, and be able to decide what physics matters most. They should become adept at seeing how physics equations and principles can give insight into what organisms do. Given the kinds of questions that biologists like to ask, you can imagine that neither frictionless vacuums nor inclined planes hold much interest.

Both of us have spent a good deal of our time in conversation with our biology colleagues, creating problems of relevance to them that are also doable by students in an introductory biology course. We give here examples of two problems, each will showcase different challenges.

The first problem is that of figuring out how big a worm can grow. This problem evolved from an interaction between a physicists (EFR) and a biologist (Cooke), trying to find matching examples that would serve to introduce students in both an IPLS class and OrgBio to the value of scaling and dimensional reasoning.[5] The original version for the biology course focused on specific numbers and particular (realistic) models of growth. The physicist's response focused on abstract symbolic relations and the expression of the results in a variety of representations. The negotiated compromise included realism, explicit discussion of modeling, mathematical abstraction, multiple representations, and functional implications. The result is shown in Sidebar 4 and has been successfully used in our IPLS class both as a homework assignment and as a group problem-solving activity.

The second problem comes from the cellular level (Sidebar 5). The students are asked to think about how gradients can cause motion, and how competing gradients can stop motion. Although the name is not mentioned, we have the beginning ingredients to derive the Nernst potential. (All we need to add the Boltzmann distribution.) What is important pedagogically is that we have left out a lot of biology: usually there is more than one mobile ion, and osmotic pressure stays at a tolerable level thanks to active pumping of ions across the cell membrane. It is both important to leave these out (to make the problem manageable) and to mention that these are left out and why. This is a golden opportunity to

**Sidebar 4:** This question was the result of negotiation between EFR and his colleague Todd Cooke.[3]

The earthworm absorbs oxygen directly through its skin. The worm does have a good circulatory system (with multiple small hearts) that brings the oxygen to all the cells. But the cells are distributed through the worm's volume and the oxygen only gets to come in through the skin -- so the surface to volume ratio plays an important role. Let's see how this works. Here are the worm's parameters.

A typical specimen of the common earthworm (*Lumbricus terrestris*) has the following average dimensions:

- Mass – 3.7 g
- Length – 12 cm
- Width – 0.64 cm

http://www.cartoonistgroup.com/store/add.php?iid=63488

The skin of the worm can absorb oxygen at a rate of $A = 0.24$ μmole (μmole = $10^{-6}$ moles) per square cm per hour. The body of the worm needs to use approximately $B = 0.98$ μmole (μmole = $10^{-6}$ moles) of oxygen per gram of worm per hour.

A. It is reasonable to model the shape of the earthworm as a solid cylinder. Using the dimensions of a typical earthworm above, calculate its surface area (ignore the surface areas of the blunt ends in all calculations), volume, and density.

B. If the worm is much longer than it is wide ($L >> R$) is it OK to ignore the end caps of the cylinder in calculating the surface area? How does the surface area and volume of the worm depend on the length of the worm, $L$, and the radius of the worm, $R$?

C. For an arbitrary worm of length $L$, radius $R$, and density $d$, write an equation (using the symbols $A$ and or $B$ rather than the numbers) that expresses the number of moles of oxygen the worm absorbs per hour and the number of moles the worm uses per hour. What is the condition that the worm takes in oxygen at a rate fast enough to survive? Does this simple model predict that the typical worm described above absorbs sufficient oxygen to survive?

D.1. Consider the effect of changing the various size parameters of a worm. First consider a worm of length 12 cm that grows by keeping its length the same but increasing its radius. Use a spreadsheet to plot the total oxygen absorbed through the skin of the worm and the total oxygen used by the worm as a function of its length from a radius of 0 cm (not really reasonable) up to a radius of 1 cm. Do the two curves cross? Explain what the crossing means and what its implications are.

D.2. Now consider a worm width 0.64 cm the grows by keeping its width the same but increasing its length. Use a spreadsheet to plot the total oxygen absorbed through the skin of the worm and the total oxygen used by the worm as a function of its length from a length of 0 cm (not really reasonable) up to a length of 50 cm. Do the two curves cross? Explain what the crossing means and what its implications are.

D.3. Write (in symbols) an equation that represents the crossover condition -- that the oxygen taken in per hour exactly equals the oxygen used per hour. Cancel common factors. Discuss how this equation tells you about what you learned about worm growth by doing the two graphs.

E. Our analysis in D was a modeling analysis. An organism like an earthworm might grow in two ways: by just getting longer or isometrically -- by scaling up all its dimensions. What can you say about the growth of an earthworm by these two methods as a result of your analysis in part D? Does a worm have a maximum size? If so, in what sense? If so, find it.

F. In typical analyses of evolution and phylogenetic histories, earthworm-like organisms are the ancestors of much larger organisms than the limit here permits. Discuss what sort of variations in the structure of an earthworm might lead to an organism that solves the problem of growing isometrically larger than the limit provided by this simple model.

talk about why simplifications are a good first step to understanding.

At UNH we surveyed our students to find the value they placed on including biological examples.[2] Eighty percent of students found the biological examples interesting, 58% found them relevant to their other courses and career plans, and 58% found that these examples helped them to understand the physics. About 15% of the students felt the biological applications did not belong in a physics class. One student who did see the value in the biology applications wrote

> *[I] strongly believe the biological applications added to this physics … course were completely relevant. Not only were the applications extremely interesting, but I think they helped students (including myself) connect adaptation abilities, and more importantly, evolutionary sequences of the natural world of organisms. Organisms understand physics incredibly well; this is how they thrive and survive. I think in order for a biology student to be successful in the future, one must make these connections at an early stage in his/her career.*

**Conclusions**

We have argued that teaching physics to biologists requires far more than making the course for engineers mathematically less rigorous and adding in a few superficial biological problems. What is needed is for physicists to work closely with biologists to learn not only what physics topics and habits of mind that are useful to biologists, but how biologists work is fundamentally different from ours, and how to bridge that gap. Fortunately, there are many educators around the country working on this challenge, so that instructors need not re-invent the wheel or go it alone. (See Sidebar 2)